\documentclass[pra,twocollumn,showpacs,amsmath,amssymb]{revtex4}
\usepackage{graphicx,color}
\usepackage{dcolumn}
\usepackage{bm}

\begin{document}

\title{The dressed atom as binary phase modulator: towards attojoule/edge optical phase-shift keying}
\author{Joseph Kerckhoff} \email{jkerc@stanford.edu}
\affiliation{Edward L.\ Ginzton Laboratory, Stanford University, Stanford, California 94305, USA}
\author{Michael A. Armen} %\email{marmen@stanford.edu}
\affiliation{Edward L.\ Ginzton Laboratory, Stanford University, Stanford, California 94305, USA}
\author{Dmitri S. Pavlichin}% \email{dmitrip@stanford.edu}
\affiliation{Edward L.\ Ginzton Laboratory, Stanford University, Stanford, California 94305, USA}
\email{dmitrip@stanford.edu}
\author{Hideo Mabuchi}% \email{hmabuchi@stanford.edu}
\affiliation{Edward L.\ Ginzton Laboratory, Stanford University, Stanford, California 94305, USA}
\email{hmabuchi@stanford.edu}

\date{\today}

\maketitle

\noindent Nanophotonic technologies offer great promise for ultra-low power optical signal processing, but relatively few nonlinear-optical phenomena have yet been explored as bases for robust digital modulation/switching~\cite{Yang07,Fara08,Liu10,Noza10}. Here we show that a single two-level system (TLS) coupled strongly to an optical resonator can impart binary phase modulation on a saturating probe beam. Our experiment relies on spontaneous emission to induce occasional transitions between positive and negative phase shifts---with each such edge corresponding to a dissipated energy of just one photon ($\approx 0.23$ aJ)---but an optical control beam could be used to trigger additional phase switching at signalling rates above this background. Although our ability to demonstrate controlled switching in our atom-based experiment is limited, we discuss prospects for exploiting analogous physics in a nanophotonic device incorporating a quantum dot as the TLS to realize deterministic binary phase modulation with control power in the aJ/edge regime.

A resonantly driven optical dipole emitter can oscillate either in-phase or completely out-of-phase with the driving field. When the emitter is modeled as a quantum TLS, the two orthogonal `dressed' states that arise in the strong driving limit~\cite{Cohe04} are accordingly coherent superpositions of the ground and excited states with $0$ or $\pi$ phase relative to the driving field. The re-radiated field of such a driven dipole has a phase corresponding to that of the dipole velocity, {\it i.e.}, $\pm\pi/2$ relative to the drive. In the setting of cavity quantum electrodynamics (cQED) with strong coupling \cite{Mabu02,Berman}, in which we consider a TLS driven by the intra-cavity field of a high-finesse optical micro-resonator, the re-radiated field can have an amplitude that is comparable to that of the driving field itself, resulting in a substantial positive or negative phase shift of the total cavity output field relative to the input~\cite{Alsi91,Armen09}. The composite cavity-TLS system thus functions as a binary optical modulator, imparting a positive phase-shift to the transmitted light when the TLS is in one dressed state and a negative phase-shift when it is in the other. Any process inducing a transition from one dressed state to the other will correspondingly change the optical phase shift of the transmitted probe beam; the underlying discreteness of the TLS Hilbert space gives rise to a {\em digital} modulation format.

A basic demonstration of binary phase modulation by a single TLS in a cavity can be performed using broadband homodyne measurement to monitor the phase of the transmitted optical field while relying on spontaneous emission of the TLS to induce random transitions between dressed states. Our experiments are performed on a standard cQED setup \cite{Mabu96,Armen09} in which a cloud of laser cooled $^{133}$Cs atoms is dropped onto a high-finesse Fabry-Perot optical resonator (length 27$\mu$m, field decay rate $\kappa/2\pi =$ 9.3MHz) with a standing wave TEM$_{00}$ mode (waist 18$\mu$m) actively stabilized at a frequency fixed relative to an atomic cycling transition (dipole decay rate $\gamma_\perp/2\pi$=2.6MHz). As individual atoms fall through the spatial profile of the cavity mode, they experience a position-dependent coupling $g(r)$ to the intra-cavity field (maximal value $g_0/2\pi$ = 56.8MHz at the cavity anti-nodes). A weak optical probe is used to detect atom transits; when an atom reaches a position of near-maximal coupling the probe strength is increased and its frequency brought close to atomic resonance in order to observe binary phase switching~\cite{Armen09}. Under these conditions the atom experiences fluctuating opto-mechanical forces that induce random variations of $g(r)$ on time scales longer than a few $\mu$s. The effects observed here nonetheless typically persist for several tens of $\mu$s, until the atom either completely exits the cavity mode or is pumped into a dark state.

\begin{figure}[tb!]
\includegraphics[width=0.85\textwidth]{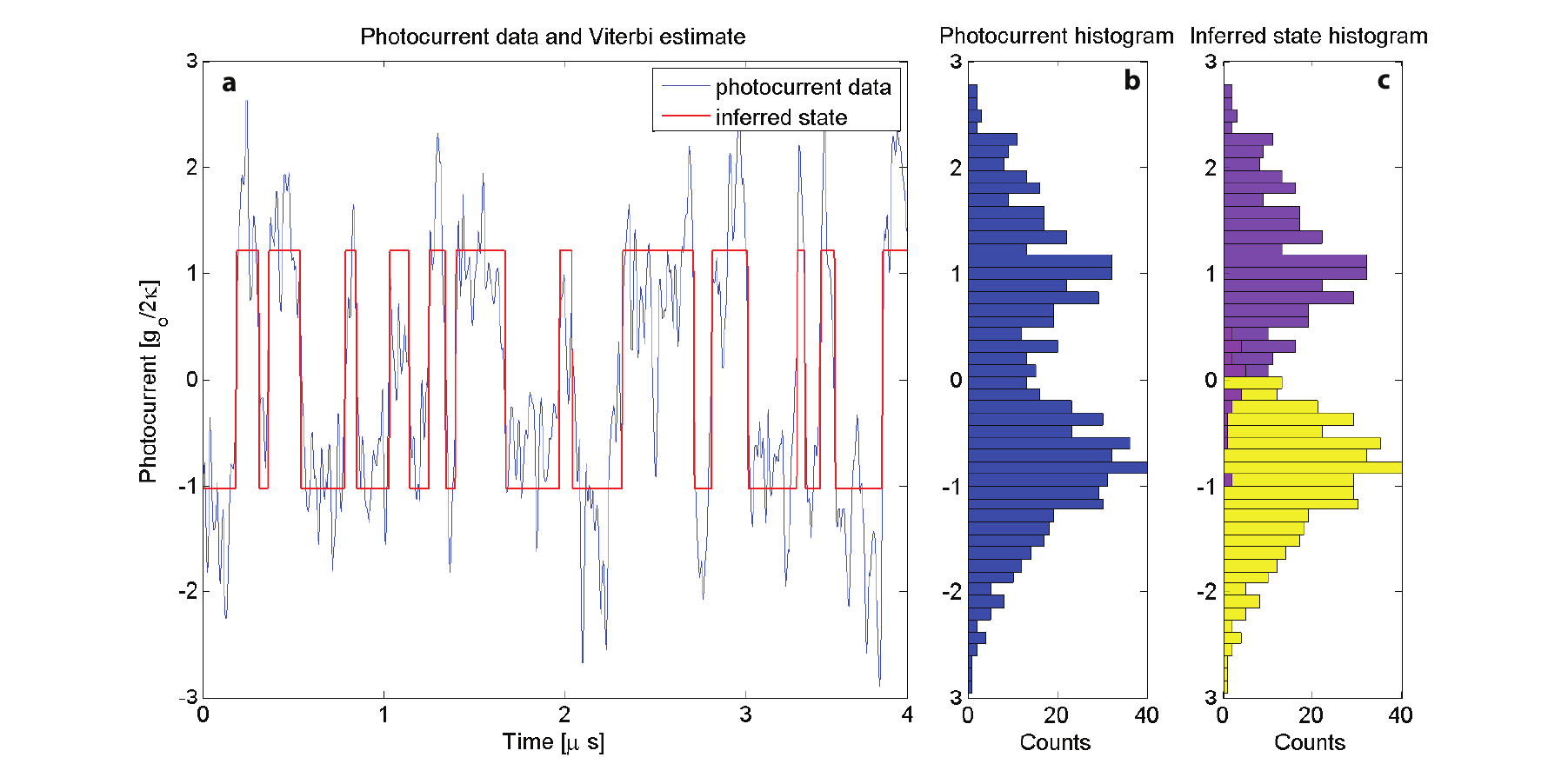}
\caption{\label{fig:Viterbi_Fit} {\bf Spontaneous binary phase modulation. a,} A representative phase-quadrature, optical homodyne measurement of the field transmitted by a resonator containing a strongly coupled TLS is displayed in blue with a 20MHz bandwidth, calibrated by the cQED parameters, according to strong-driving, theoretical predictions. Red overlay is the decoded binary signal produced by the Viterbi algorithm with HMM parameters obtained via EM (see text). {\bf b}, A histogram of the photocurrent data segment in {\bf a}, displaying a dual-Gaussian distribution consistent with binary phase modulation and with theoretical predictions. {\bf c}, Dual histograms of the photocurrent data, each taking counts only when the Viterbi path occupies the positive or negative state.}
\vspace{-0.1in}
\end{figure}

To minimize opto-mechanical effects we work with barely-saturating probe intensity and detune the frequencies of the probe and cavity mode $\sim\kappa$ below atomic resonance (resulting in optical dipole forces towards the cavity anti-nodes). Figure~\ref{fig:Viterbi_Fit}a depicts an exemplar phase-quadrature homodyne measurement taken during an atom transit with 9MHz detuning and 380pW probe power (which would produce a $\langle \hat{n}\rangle=14$ photon coherent state in the cavity if there were no atom), and $\eta=.20$ overall detection efficiency. The photocurrent segment (recorded at 200MS/s but displayed with 20MHz bandwidth for clarity) resembles a random telegraph with added Gaussian noise, featuring sporadic jumps between levels indicating positive and negative phase shifts of the transmitted optical beam. The photocurrent histogram in Figure~\ref{fig:Viterbi_Fit}b suggests a dual-Gaussian distribution in this single shot segment, in agreement with theoretical expectations. Although a complex physical model is required to describe the TLS-cavity system, the output of a device that is claimed to function as a random binary phase modulator should be well described by a two-state, time-invariant hidden Markov model (HMM)~\cite{HMM}. Allowing for unequal transition rates between the two HMM states and assuming only that each state produces normally-distributed (but otherwise unspecified) photocurrent, we apply a standard expectation maximization (EM) algorithm \cite{Welch03,HMM} to full-bandwidth photocurrent data segments to find the HMM parameters that best fit the data. The red trace overlay in Figure~\ref{fig:Viterbi_Fit}a represents the decoded binary signal produced by the Viterbi algorithm~\cite{Viterbi,HMM} using optimal parameters from the EM procedure. The histograms in Figure~\ref{fig:Viterbi_Fit}c were constructed by segregating the photocurrent according the Viterbi path, which apparently results in two near-normal distributions. We note that these conditional distributions mainly reflect optical shot-noise but are broadened by random variation of $g(r)$.

\begin{figure}[tb!]
\includegraphics[width=0.85\textwidth]{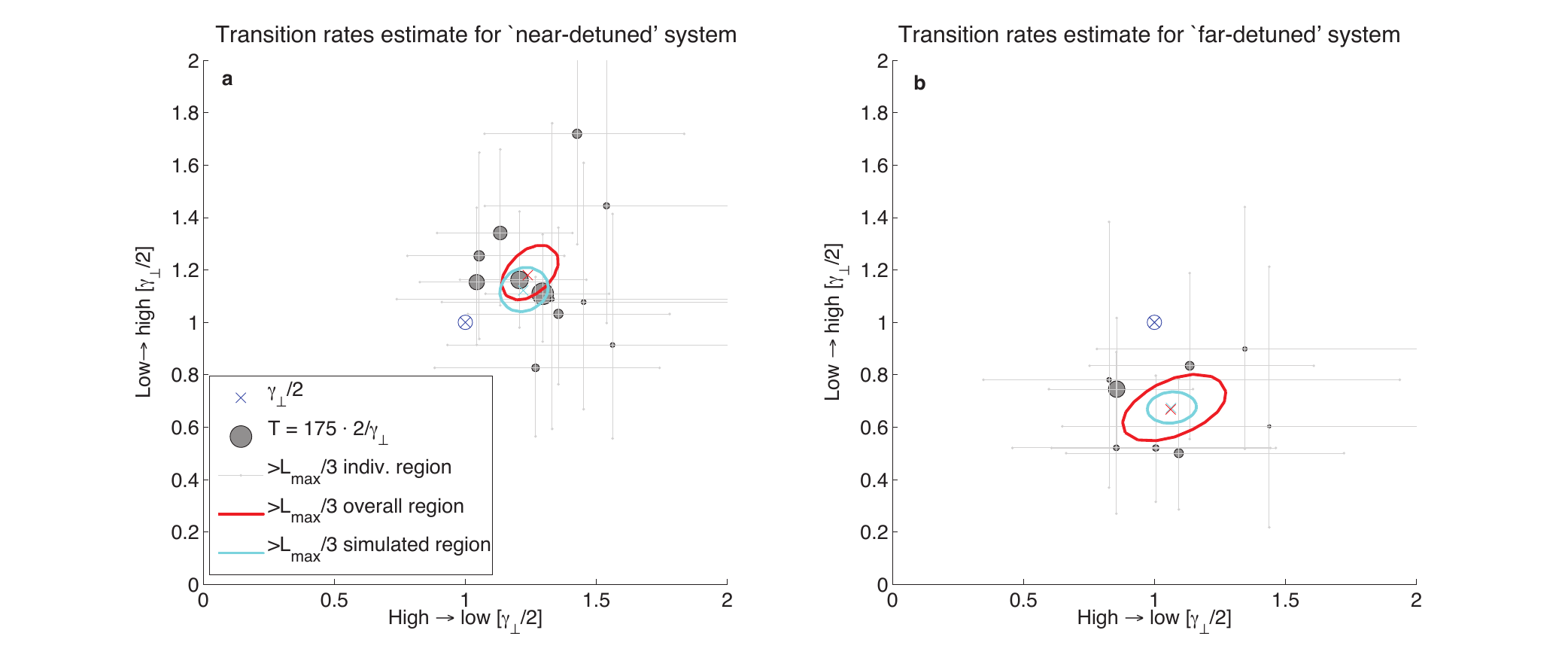}
\caption{\label{fig:RatesEstimate} {\bf Estimation of spontaneous transition rates.} Both plots depict contours of the likelihood functions of HMM transition rates for low$\leftrightarrow$high switching. The most likely rate-pairs inferred from data taken from individual atom transit segments are indicated by grey dots, with the diameter of each dot representing the duration of the corresponding data segment. The grey bars represent the intervals over which the likelihood of the individual transits' rate-pairs are at least 1/3 of their maximum. The red cross locates the most likely rate-pair, given an aggregated segment formed from the individual transit measurements, with the red oval enclosing the region of rate-pair likelihoods that are at least 1/3 of the maximum. The teal cross and oval represent the same likelihood contours produced from simulated data. {\bf a}, Likelihood contours using a `near-detuned' system, with atomic detuning $\Delta/2\pi = 9$MHz, and a probe strength corresponding to $\langle \hat n\rangle=14$ photons in an empty cavity, for which a symmetric switching rate-pair slightly greater than $\gamma_\perp/2$ is expected. {\bf b}, The same as {\bf a}, except using a `far-detuned' system with $\Delta/2\pi=40$MHz, $\langle \hat n\rangle=19$, for which an asymmetric switching model is anticipated.}
\vspace{-0.1in}
\end{figure}

Binary decoding of the experimental photocurrent segments yields transition rate estimates that match predictions of a detailed physical model. The plots in Figure~\ref{fig:RatesEstimate} depict contours of the likelihood functions of HMM parameters estimated from individual and aggregated atom transits, obtained under two different input conditions. Figure~\ref{fig:RatesEstimate}a reflects data acquired from a `near-detuned' system with the same parameters as in Figure~\ref{fig:Viterbi_Fit}. The high likelihood of a symmetric switching model apparent here is both theoretically predicted and consistent with the symmetric photocurrent histograms in Figure~\ref{fig:Viterbi_Fit}. The slightly greater than $\gamma_\perp/2$ switching rates of the most likely models (as opposed to exactly $\gamma_\perp/2$, which occurs in the strong-driving limit) is due to the moderate power of the probe. Figure~\ref{fig:RatesEstimate}b reflects data taken with cavity and probe held 40MHz below atomic resonance. A significant asymmetry in the transition rates is expected and inferred from the likelihood contours. Both figures compare parameter estimates obtained from experimental data to estimates from simulated photocurrent data of comparable aggregate duration and detection efficiency (quantum trajectory simulations of the driven damped Jaynes-Cummings Master Equation~\cite{Tan99,Carm93} assuming no atomic motion and a true two-level atom). For both detunings the likelihood contours for the experimental and simulated signals largely overlap, indicating that our fundamental device physics model accounts well for the observed binary switching statistics.

\begin{figure}[tb!]
\includegraphics[width=0.65\textwidth]{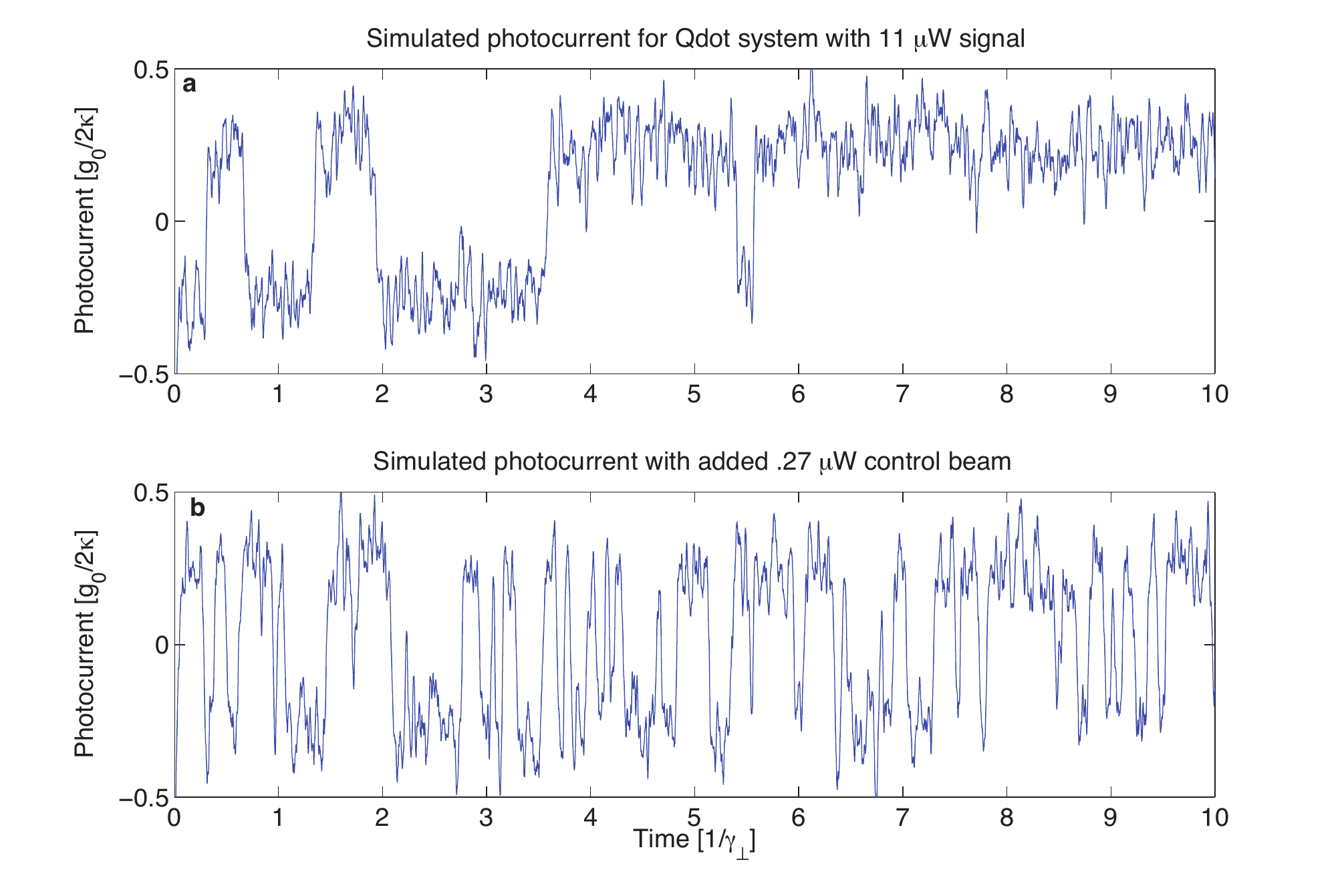}
\caption{\label{fig:Qdot} {\bf Simulated controlled switching in a Qdot-PC cQED system. a}, Simulated photocurrent segment assuming cQED parameters demonstrated in \cite{Faraon10}, a 11$\mu$W probe and perfect detection efficiency at a 10GHz detection bandwidth. The time axis is calibrated in terms of the Qdot dephasing rate of $\gamma_\perp/2\pi = 0.1$GHz and the photocurrent axis is calibrated by the ratio $g_0/\kappa$ ($\{g_0,\kappa\}/2\pi=\{20,40\}$GHz in \cite{Faraon10}).
{\bf b}, Simulated photocurrent segment assuming the same system, plus a .27$\mu$W cw `control' beam detuned by the magnitude of the TLS-field coupling Hamiltonian, {\it i.e.}, by -400GHz relative to the probe beam. Even this weak control probe induces an order of magnitude more flops than the intrinsic spontaneous emission dynamics, corresponding to less than 100aJ dissipated by the control probe per induced edge.}
\vspace{-0.1in}
\end{figure}

Although the signals discussed above reflect only random switching induced by spontaneous emission, external optical control of the dressed-TLS switch state is possible. One intuitive strategy (considering the elementary dressed-state model~\cite{Cohe04}) would be to drive the system with an additional `control' beam that is detuned from the `signal' beam by a frequency corresponding to the expected magnitude of the atom-field coupling. A constant control beam of this type should simply increase the rate of random phase switching, but deterministic switching could be achieved in principle by modulating the control beam power using real-time feedback (of course, with error correcting codes mere modulation of the switching rate could suffice for communication). Such an approach, although intuitive, is sub-optimal as it attempts to drive coherent transitions between dressed states and is thus in direct competition with dissipative dynamics that enforce the digitization of the transmitted signal. Nonetheless, it is possible to find candidate cQED systems that should exhibit both good binary switching and controllability via this simple method. Such systems appear when $g/\kappa$ is $\sim 1$ and the TLS dephasing rate is relatively slow, as can occur in solid-state nanophotonic systems (which are also attractive for integrability and GHz dynamics). In Figure~\ref{fig:Qdot} we present simulations of switching-rate control using parameters from a demonstrated quantum dot, photonic crystal cavity \cite{Faraon10}, assuming perfect detection efficiency. The addition of an appropriately detuned, cw control beam with $<5\%$ of the signal power is able to induce switching dynamics an order of magnitude faster than the spontaneous emission background rate. This control beam power corresponds to less than 100aJ per induced phase-switch. Although quite low, this switching energy is not yet at the presumed limit of a single dissipated photon (sub-aJ per edge) exemplified by the spontaneous emission-driven random switching. Control-pulse shaping, coherent feedback and/or reservoir engineering could further reduce the power required for fast deterministic phase modulation.

\begin{figure}[tb!]
\includegraphics[width=0.75\textwidth]{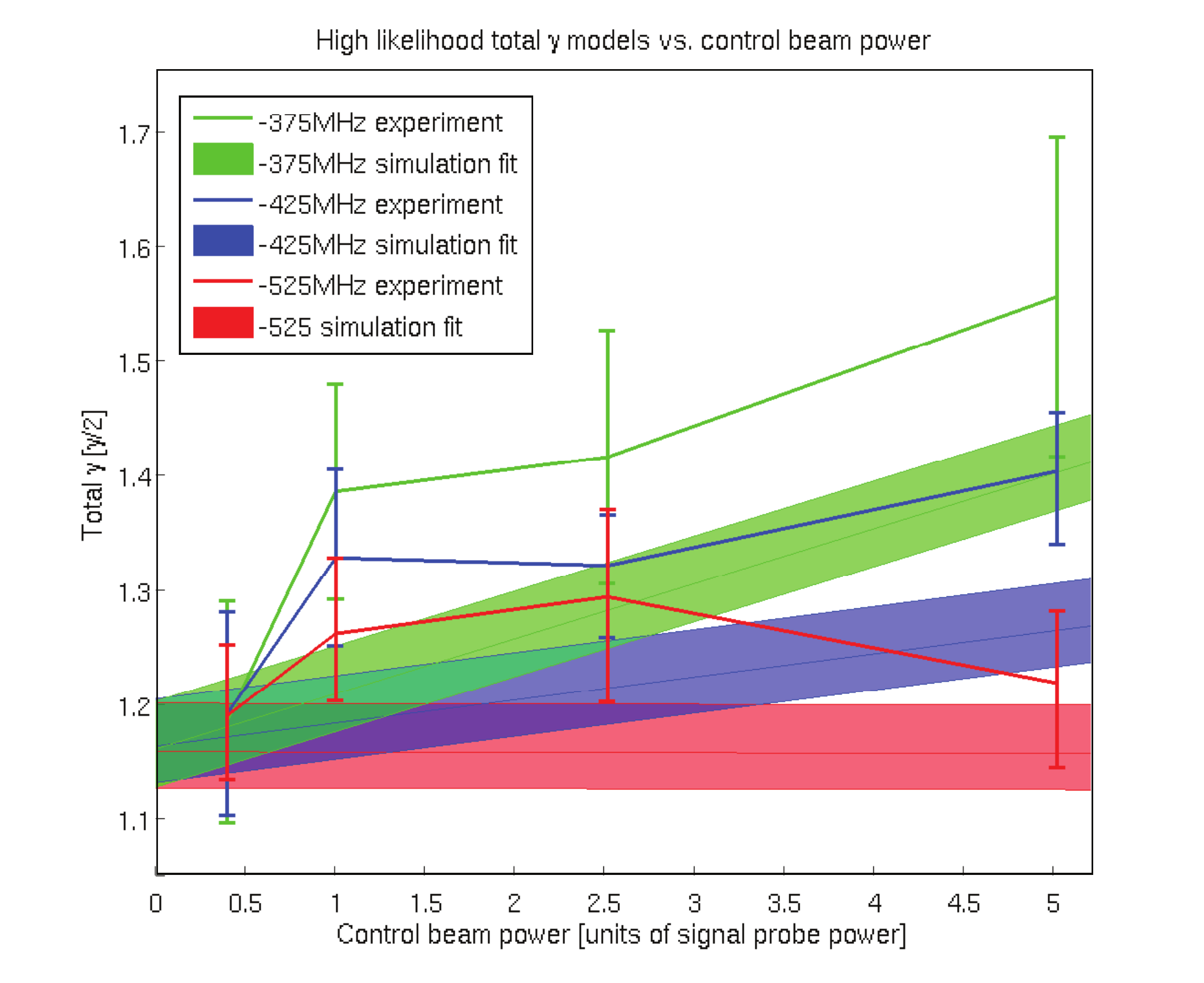}
\caption{\label{fig:g_vs_Pwr} {\bf Degree of controllability observed in the atomic system.} Data points represent inferred `total $\gamma$' (mean rate at which transitions in either direction occur) for various control beam parameters. The thin lines track the most likely total $\gamma$ and the confidence interval represents the range of $\gamma$ values with at least 1/3 the maximum likelihood. The green, blue and red data points depict the total $\gamma$ as a function of control power for a control probe detuned by -375MHz, -425MHz, and -525MHz from the signal probe, respectively. The colored shaded regions represent linear best fits of the corresponding confidence intervals based on a simulated data set with perfect detection efficiency and duration several times that of the experimental data.  The apparent positive, constant bias in the experimental $\gamma$ relative to simulation for control powers at least equal to the probe's is perhaps due to enhanced opto-mechanical motion induced by a strong control beam.  Although the effect of the control is marginal in both experiment and simulation, as anticipated for this atomic system, the results are consistent with the basic concept of external, optical controllability explained in the text and illustrated more convincingly in a Qdot-PC simulation in Figure~\ref{fig:Qdot}. }
\vspace{-0.1in}
\end{figure}

Despite our atomic system's large $g/\kappa$ ratio, which is unfavorable for the control method simulated in Figure~\ref{fig:Qdot}, we have attempted to demonstrate stimulated switching experimentally using the `near-detuned' system. Figure~\ref{fig:g_vs_Pwr} shows inferred `total' transition rates (the mean rate at which switching in either direction occurs) as a function of the power of the control beam, for three different control beam frequencies. In the strong driving limit the optimum control detuning would be 425MHz, but given our moderate signal power the effective coupling energy (and therefore optimal control detuning) should be somewhat smaller. The statistical significance of the data displayed in Figure~\ref{fig:g_vs_Pwr} is marginal, but the data do show consistent trends that match expectations: data taken with -375MHz and -425MHz control detuning induce higher transition rates with higher power, with -375MHz consistently higher, while the -525MHz detuning induced essentially no net transition increase over the accessible range of control powers (limited by opto-mechanical effects on the high end). Linear fits to simulations reproduce these general trends of the data and confirm the limited degree of control achievable in our experimental system. We thus infer an apparent control energy requirement of only $\sim 1$fJ/edge even in our atomic implementation (with unfavorable parameters for efficient control) of the dressed-state binary phase modulator.

In conclusion, we have demonstrated that a TLS-cavity system can impart controllable binary phase modulation on a saturating optical probe. The fundamental device physics indicate that control energy efficiencies in the aJ/edge regime should be achievable with current capabilities in nanophotonics.

\section{methods}
\subsection{Experimental setup}
The experiment consists of a standard cQED setup~\cite{Armen09,Mabu96} utilizing laser cooled $^{133}$Cs atoms and a high-finesse Fabry-Perot optical resonator. We attempt to drive only the $(6S_{1/2},F=4,m_F=+4)\rightarrow(6P_{3/2},F=5,m_F=+5)$ atomic cycling transition at 852nm through frequency- and polarization-selectivity so that the atom may be approximated as a TLS. The use of an improved cavity geometry and mirror mounting scheme, as compared to a previous experimental study of single-atom cavity QED in the strong driving regime~\cite{Armen09}, was crucial in enabling us to observe clear binary phase modulation.

Inside a UHV ($\approx10^{-9}$Torr) chamber and placed on a multi-stage vibration-isolation stack, the Fabry-Perot optical resonator is formed by two high-reflectivity (8ppm transmission, 2ppm loss), 10cm radii of curvature dielectric mirrors with roughly 27$\mu$m of separation, yielding a 300,000-finesse optical resonator for the standing wave, TEM$_{00}$, 18$\mu$m-waist transverse spacial mode with a field decay rate of $\kappa = 2\pi\times9.3$MHz. We took particular care to mount the mirrors in a rotationally-symmetric manner to minimize stress-induced birefringence in the mirror coatings, allowing for full polarization-selectivity of the atomic transitions. The cavity length is tuned and actively stabilized by two shear-mode piezoelectric plates underlying the two mirror mounts. The precise cavity length and resonance frequency is continually stabilized by the Pound-Drever-Hall~\cite{PDH} method using an additional laser probe detuned by the desired probe/cavity resonance frequency by two cavity free spectral ranges (at an optical wavelength of roughly 826nm, which interacts negligibly with Cs).

A Doppler-limited, magneto-optically trapped ensemble of $\sim10^6$ atoms is formed roughly 1cm above the cavity mode in the UHV chamber. After cooling, the ensemble trap is switched off, allowing the cold atoms to fall under gravity towards the cavity mode and by the time they reach the cavity mode their free-fall velocity tends to dominate any residual thermal motion. Due to the strong coupling between the targeted atomic transition and the cavity mode (with calculated maximum value max$_r\{g(r)\}\equiv g_0=2\pi\times56.8$MHz at the cavity anti-nodes, using the dipole strength of the atomic transition and cavity mode volume), individual atom transits are detected by monitoring the ($g(r)$-dependent) cavity transmission amplitude using a relatively weak and near-resonant probe, a free space balanced photodetector, and an actively phase-locked optical local oscillator. Once a near-maximally coupled atom has been detected, the probe power and frequency shift to the desired experimental levels and data acquisition is initiated.  Although multiple atom transits per drop may be visible, the atomic ensemble is sufficiently diffuse such that no more than one atom is simultaneously present in the cavity mode and we acquire data from only one transit per ensemble drop. Due to the many sensitive stabilization requirements and slow drifts in the experimental apparatus, data are usually analyzed in groups of 50 atom transits, over which experimental stability can be confidently maintained.

Within a group of 50 transits records, photocurrent segments are selected for analysis in a two-stage process. First, photocurrent segments in each transit with above-shot noise variance (corresponding to TLS-induced binary phase switching) are algorithmically identified using HMM methods. Typically these segments persist for several tens of microseconds. Next, some subjective selection of these algorithmically-identified segments is required to limit the analysis to segments over which the variance is both reasonably high and constant in time, corresponding to switching signals in which the TLS maintained near-maximal coupling throughout. Independent trials of this selection process were found to result in similar final results. Despite this attempt to compensate for the time-dependent atom-field coupling (due to a position-dependent $g(r)$), modulation in the switching variance is typically apparent over timescales greater than a few microseconds; in fact, a significant fraction of the transit segments display near-sinusoidal modulation in the switching variance, corresponding to atomic motion through several standing wave anti-nodes.
%Although many 50-transit iterations were typically made for each experimental configuration, comparisons between configurations are typically made on this convenient, uniform basis.  For instance, Figure \ref{fig:RatesEstimate} compares the analysis of selected photocurrent segments from two 50-transit data sets for two different experimental configurations.

\subsection{The Jaynes-Cummings model and simulation}
The Jaynes-Cummings Hamiltonian~\cite{Berman,JC} is widely used to describe the internal dynamics of a cQED system driven by a coherent probe (using $\hbar=1$):
\begin{equation}
H = \Delta \sigma^\dag\sigma+\Theta a^\dag a + ig(r)(a^\dag \sigma-a\sigma^\dag) + i\mathcal{E}(a^\dag-a)
\end{equation}
where $a$ is the annihilation operator for the cavity mode, $\sigma = \vert g\rangle\langle e\vert$ is the TLS lowering operator and $^\dag$ signifies the Hermitian conjugate.  $\Delta$ is the detuning between the probe and the atomic transition frequencies and $\Theta$ is the detuning between the cavity mode resonance and the probe ($\Theta=0$ in all systems considered here).  The third, `coupling' term represents the interaction between the atomic transition and cavity mode and describes the process by which quanta of energy are exchanged (at rate $\propto g(r)$) between the TLS and mode.  The final term represents the coherent driving term, with amplitude $\mathcal{E}$.

The physical damping processes included in the overall dynamics model include the decay of cavity photons out of the resonator at a rate of $2\kappa$ per intra-cavity photon and the spontaneous emission of an excited TLS at a rate of $2\gamma_\perp$.  Standard quantum trajectory theory~\cite{Carm93} based on these processes underlies much of the theoretical analysis and numerical simulation.  For example, simulation of a photocurrent given a set of experimental parameters first involves the calculation of a possible trajectory for the internal quantum state vector of the cQED system $\vert\psi_c(t)\rangle$ by numerically integrating the stochastic Schr\"{o}dinger equation \cite{Tan99,Carm93}
\begin{eqnarray}
d\vert\psi_c(t)\rangle &=& -(iH+\kappa a^\dag a+\gamma_\perp\sigma^\dag\sigma)\vert\psi_c(t)\rangle dt+\left(\sqrt{2\kappa}\langle\psi_c(t)\vert a+a^\dag\vert\psi_c(t)\rangle dt+dW_t^{(1)}\right)\sqrt{2\kappa}a\vert\psi_c(t)\rangle+\nonumber\\
&&\left(\sqrt{2\gamma_\perp}\langle\psi_c(t)\vert \sigma+\sigma^\dag\vert\psi_c(t)\rangle dt+dW_t^{(2)}\right)\sqrt{2\gamma_\perp}\sigma\vert\psi_c(t)\rangle
\end{eqnarray}
where $\{dW_t^{(1)},dW_t^{(2)}\}$ are randomly generated, independent Wiener increments and the state vector is forcibly re-normalized after each recursive update.  The simulated photocurrent $dQ(t)$ may then be calculated using this state trajectory and calibrated detection efficiency $\eta$
by
\begin{equation}
dQ(t) = \sqrt{\eta}\left(\sqrt{2\kappa}\langle\psi_c(t)\vert a+a^\dag\vert\psi_c(t)\rangle dt+dW_t^{(1)}\right)+\sqrt{1-\eta}dW_t^{(3)}
\end{equation}
where $dW_t^{(3)}$ is a third, independent Wiener increment.  Although this model may be generalized to include a time dependence in the coupling rate $g(r)$ and a more realistic, multi-level atomic structure, all simulations here utilized the maximal $g_0$ for the static coupling rate and assumed a TLS atomic model.

\subsection{Hidden Markov model analysis}
A hidden Markov model (HMM) \cite{HMM} is a common method for analyzing systems with a dynamically evolving state that is monitored through noisy observations.  The model is Markovian in the sense that the probability that the system will be in a particular state in the next time step is a function only of its current state. Our HMM analysis of experimental and simulated photocurrents assumes only two possible internal states (each corresponding to one of the two atomic dressed states), with transitions between them at fixed (and generally asymmetric) rates. Inference of the `hidden' internal state trajectory is made using the photocurrent measurements and models for the mean transition rates and measurement distributions associated with each internal state. We expect these photocurrent distributions to be normal, with the same variance (reflecting optical shot noise as well as $g(r)$ variations), but differing means (corresponding to the associated positive and negative phase shifts).

The two state HMM with normally-distributed photocurrent `emissions' most likely to produce a particular photocurrent segment ({\it i.e.} the maximum likelihood estimator for the HMM) is calculated according to the Baum-Welch expectation maximization algorithm \cite{Welch03,HMM}.  Essentially, this algorithm consists of first assuming some set of HMM parameters: a pair of state transition rates and a mean and variance of the emissions distributions associated with each state.  Then, for the particular photocurrent segment, an estimate of the hidden state trajectory is made and, using this estimate, the segment-average transition rates and emission statistics are calculated.  It can be shown that these inferred, average HMM parameters comprise a model with a necessarily higher likelihood than the originally assumed one.  Thus, the procedure is iterated, each time using the inferred, segment-averaged HMM parameters from the previous trajectory estimate until the model parameters converge.  Once this maximum likelihood HMM is identified, the likelihoods of models in the vicinity of the most likely model are also calculated.  The Viterbi state trajectory estimate (such as used in Figure \ref{fig:Viterbi_Fit}) maximizes the probability of the entire state trajectory, given a HMM and a photocurrent segment \cite{Viterbi,HMM}.

\begin{acknowledgments}
JK would like to thank Arka Majumdar and Ramon van Handel for helpful discussions.  This work is supported by DARPA-MTO under Award No. FA8650-10-1-7007.  DSP acknowledges the support of a Stanford Graduate Fellowship. 
\end{acknowledgments}

\section{Additional information}
The authors declare no competing financial interests.

\end{document}